# 2-Dimensional Tunnel FETs with a Stable Charge-Transfer-Type p$^+$-WSe$_2$ Source

*Junyang He, Nan Fang, Keigo Nakamura, Keiji Ueno, Takashi Taniguchi, Kenji Watanabe, and Kosuke Nagashio*[*]

Junyang He, Nan Fang, Keigo Nakamura, and Prof. Kosuke Nagashio[*]
Department of Materials Engineering
The University of Tokyo
Tokyo 113-8656, Japan
[*]nagashio@material.t.u-tokyo.ac.jp
Prof. Keiji Ueno
Department of Chemistry
Saitama University
Saitama 338-8570, Japan
Dr. Takashi Taniguchi, Dr. Kenji Watanabe
National Institute of Materials Science
Ibaraki 305-0044, Japan

**ABSTRACT:**
Two-dimensional (2D) materials are highly promising for tunnel field effect transistors (TFETs) with low subthreshold swing and high drive current because the shorter tunnel distance and strong gate controllability can be expected from the van der Waals gap distance and the atomically sharp heterointerface formed independently of lattice matching. However, the common problem for 2D-2D TFETs is the lack of highly doped 2D materials with the high process stability as the sources. In this study, we have found that *p*$^+$-WSe$_2$ doped by charge transfer from a WO$_x$ surface oxide layer can be stabilized by transferring it onto a *h*-BN substrate. Using this *p*$^+$-WSe$_2$ as a source, we fabricate all-solid-state 2D-2D heterostructure TFETs with an Al$_2$O$_3$ top gate insulator, i.e., type-II *p*$^+$-WSe$_2$/MoS$_2$ and type-III *p*$^+$-WSe$_2$/WSe$_2$. The band-to-band tunneling and negative differential resistance trends are clearly demonstrated at low temperatures. This work suggests that high doped 2D crystal of the charge transfer type is an excellent choice as sources for TFETs.

## 1. Introduction

Tunnel field effect transistors (TFETs)[1-3] are promising for low power switching in digital logic circuits to supplement the conventional metal-oxide-semiconductor FET because of its potential to reduce the power dissipation by reducing the power supply voltage. This potential results from the ability of the TFET to achieve a subthreshold swing (SS) of less than 60 mVdec$^{-1}$ at room temperature (RT) because the high-energy tail of the Boltzmann distribution of carriers in the source are cut off by the band gap. Although SSs of less than 60 mVdec$^{-1}$ have been reported for various materials including CNTs,[4] Si,[5] SiGe,[6] and III-V,[7] they continue to suffer from a low drive current, which can be expected from the tunneling process. To overcome this issue and improve the performance of TFETs, the reduction of tunneling distance and tunneling barrier height as well as the efficient gate controllability for the semiconductor channel are crucial.

In these regards, the recent simulations[8-10] predict that two-dimensional (2D) materials are highly promising as TFETs with both low SS and high drive current because the shorter tunnel distance and strong gate controllability can be expected from the van der Waals gap distance, and the atomically sharp heterointerface formed independently of lattice matching and the ideally dangling bond-free layered surface. Although many 2D-2D systems[11-16] such as MoS$_2$/WSe$_2$,[12] SnSe$_2$/WSe$_2$,[13] MoS$_2$/black phosphorus (BP),[15] WSe$_2$/SnSe$_2$,[16] and BP/ReS$_2$[17] have been investigated since the first experimental demonstration of band-to-band tunneling (BTBT) in an MoS$_2$/WSe$_2$ diode,[11] the expectedly low SS using 2D-2D tunneling has not been realized. The dominant common problem is the lack of highly doped 2D materials with air stability as the sources, although there are many other issues to solve because of an early stage of 2D-TFET studies. Both BP and SnSe$_2$ used as the *p*$^+$-source are notably unstable in air, which deteriorates the interface.[18,19] Therefore, the significantly low SS of 31.1 mVdec$^{-1}$ for four orders has only been obtained from a conventional highly doped germanium source with a MoS$_2$ channel by using a liquid ion gate with extremely high capacitance.[20] This unfavorable circumstance for source selection in 2D-2D heterostructures results from the challenging situation that substitutional doping by the conventional ion implantation technique is not suitable for 2D materials because of defect formation.[21,22]

Recently, a strong $p^+$-WSe$_2$ has been demonstrated by charge-transfer doping from the self-limiting WO$_x$ surface oxide layer, which is formed by an ozone treatment.[23, 24] If this charge-transfer-type $p^+$-WSe$_2$ is used as the source in the 2D-2D TFET, the atomically sharp band alignment at the $p^+$-2D/$n$-2D interface will be formed because the charge transfer occurs at the $p^+$-2D/$n$-2D interface to cancel the energy difference in Fermi level ($E_F$), and the screening length determines the sharpness of the band alignment.[25] This procedure differs from the diffuse band alignment for the conventional depletion type interface, where substitutional ions are fixed in the lattice. Although the air exposure of $p^+$-WSe$_2$ reduces the hole doping concentration, we have found that $p^+$-WSe$_2$ can be stabilized in air by transferring it onto $h$-BN. In this study, using ozone-treated $p^+$-WSe$_2$ as a source, we fabricate all-solid-state 2D-2D heterostructure TFETs with an Al$_2$O$_3$ top gate insulator, i.e., type-II $p^+$-WSe$_2$/MoS$_2$ and type-III $p^+$-WSe$_2$/WSe$_2$, and demonstrate BTBT and a negative differential resistance (NDR) trend at low temperature.

## 2. Results and Discussion
### 2.1. Stable $p^+$-WSe$_2$ formation

In terms of the source in TFET, the multilayer $p^+$-WSe$_2$ (more than 10 layers) is more suitable than the monolayer because of the larger density of states ($DOS$) and small energy gap ($E_G$) of ~1 eV.[26] First, based on the previous experiment,[23] the O$_3$ annealing condition to form WO$_x$ on the multilayer WSe$_2$ prepared from the exfoliation of bulk WSe$_2$ crystals was optimized as follows: temperature, 200°C; annealing time, 1 hour; and O$_3$ concentration, 2800 ppm. For this condition, ~6 layers of WSe$_2$ are oxidized to WO$_x$. The detailed experiment is provided in **Figure S1** (Supporting Information).

WO$_x$ on WSe$_2$ takes electrons from WSe$_2$ because of the high electron affinity of WO$_x$ and results in $p^+$-WSe$_2$. However, in the device fabrication process, $p^+$-WSe$_2$ becomes $p$-WSe$_2$. This is not due to the contact issue related with the WO$_x$ layer at the metal/WSe$_2$ interface, because the WO$_x$ layer is reported to reduce the contact resistance by reducing the Schottky barrier.[24] Therefore, it has been considered that the electron affinity change of WO$_x$ by the air exposure causes the recovery from $p^+$-WSe$_2$ to $p$-WSe$_2$.[24] Therefore, to stabilize $p^+$-WSe$_2$, the WO$_x$ surface is transferred onto $h$-BN crystals, as shown in **Figure 1**a. After the O$_3$ treatment of WSe$_2$ on an SiO$_2$/Si substrate, WO$_x$/WSe$_2$ is picked up by a polydimethylsiloxane (PDMS) sheet. This WO$_x$/WSe$_2$ on PDMS-1 is again transferred to PDMS-2, where the WSe$_2$ surface contacts PDMS-2. Finally, the WO$_x$ layer is encapsulated by $h$-BN and WSe$_2$ after the WO$_x$ surface is transferred onto $h$-BN using an alignment system.[27] **Figure 1**b shows the transfer characteristics of two types of back-gate-type WSe$_2$ FETs, where $p^+$-type transfer characteristics are initially obtained for both cases. The Ni electrode is used for the contact. Although WSe$_2$ FET with the top WO$_x$ layer exhibits $p$-type characteristics after 1 hour in air, WSe$_2$ FET with the WO$_x$ layer encapsulated by $h$-BN and WSe$_2$ continues showing degenerate transfer characteristics even after 1 week of exposure in air. Moreover, even after the Al$_2$O$_3$ top-gate fabrication by atomic layer deposition (ALD),[28,29] the degenerate transfer characteristics with negligible temperature dependence are clearly observed for $p^+$-WSe$_2$ FET, as shown in **Figure 1**c. This behavior is notably important because the temperature dependence of the transport properties at the 2D-2D heterointerface in TFET can be discussed without being blinded by the source properties.

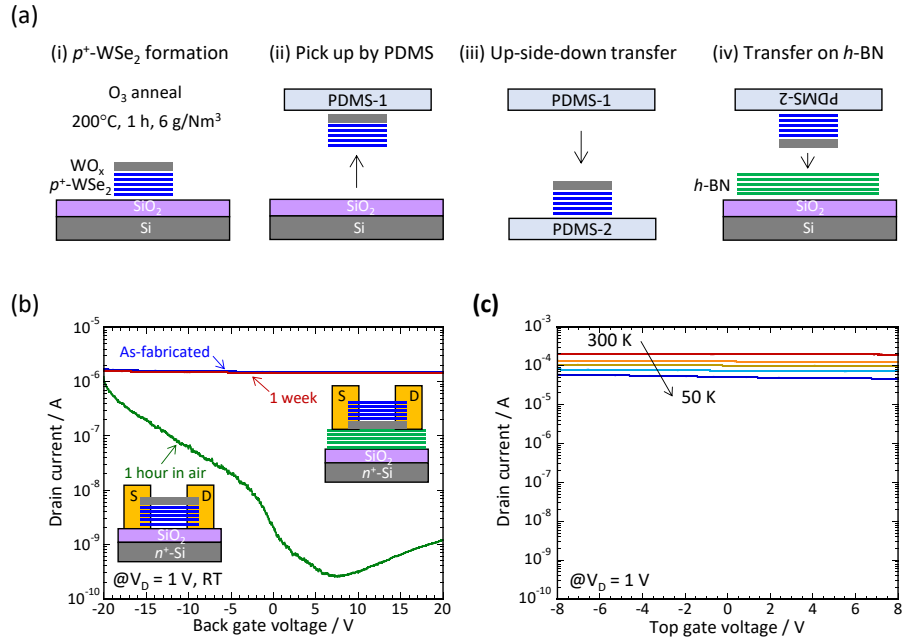

**Figure 1**. a) Schematic drawing for the procedures of (i)~(iv) to stabilize $p^+$-WSe$_2$. b) Transfer characteristics of two types of back gate WSe$_2$ FETs at RT. c) Transfer characteristics of stabilized $p^+$-WSe$_2$ FET with Al$_2$O$_3$ top gate (~30 nm), showing negligible temperature dependence (300, 200, 150, 100, and 50 K).

**2.2 $p^+$-WSe$_2$/$n$-MoS$_2$ type-II heterostructure**

The $p^+$-WSe$_2$/$n$-MoS$_2$ heterostructure on the $h$-BN substrate was fabricated using the dry transfer method with PDMS in the alignment system. The heterostructure is more easily fabricated on the $h$-BN substrate than on the SiO$_2$/Si substrate because the hydrophobic surfaces of the 2D crystals strongly adhere to each other in addition to the atomically flat surface of $h$-BN. **Figure 2** shows optical micrograph and schematic drawing of the $p^+$-WSe$_2$/$n$-MoS$_2$ heterostructure TFET with an ALD-Al$_2$O$_3$ insulator with a thickness of ~60 nm. The thickness of $p^+$-WSe$_2$ with the WO$_x$ layer, MoS$_2$, and $h$-BN is ~25 nm, ~3 nm (~4-5 layers), and ~50 nm, respectively. The heterointerface area is ~13 μm$^2$. In terms of contact metal, Ni was selected for both $p^+$-WSe$_2$ and $n$-MoS$_2$ because Ni works as the hole injection metal for $p^+$-WSe$_2$ (**Figure S2**, Supplemental Information) and electron injection metal for $n$-MoS$_2$.[30] The single-electrode metal is advantageous to reduce the device fabrication process.[11]

The electrical properties of fabricated devices are measured in a vacuum prober at different temperatures. To fully discuss the diode properties in the $p^+$-WSe$_2$/$n$-MoS$_2$ heterostructure, the transfer characteristics of the multilayer MoS$_2$ FET with the Al$_2$O$_3$ top gate, which is fabricated at a different position on the same SiO$_2$/Si wafer, are first investigated. **Figure S3**a (Supporting Information) shows the typical $n$-type conduction and negligible temperature dependence of the drain current in the high-gate-bias region even for the top-gate modulation, which suggests that the temperature dependence of the drain current due to the multilayer MoS$_2$/Ni Schottky contact is negligible. Now, let us compare the drain current as a function of the drain voltage for the $p^+$-WSe$_2$/$n$-MoS$_2$ heterostructure at 300 K and 20 K, as show in **Figure S3**b,c (Supporting Information). This drastic decrease in drain current mainly results from the suppression of the thermally excited carrier through the $p^+$-$n$ junction, since the temperature dependence of the drain current because of both Ni/$p^+$-WSe$_2$ and Ni/$n$-MoS$_2$ contacts are negligible.

Therefore, first, the diode properties in the $p^+$-WSe$_2$/$n$-MoS$_2$ heterostructure at 20 K are shown as a function of the top-gate voltage ($V_{TG}$) without the back-gate voltage ($V_{BG}$ = 0 V) in **Figure 3**a. The forward bias is placed on the positive side (the drain bias is indeed negative), and vice versa for the reverse bias (the drain bias is positive). At $V_{TG}$ = -14 V, MoS$_2$ is completely off state, and there is no current for the entire bias range. For the range of (i) -14 V < $V_{TG}$ < -9 V (representative data; $V_{TG}$ = -11 V), MoS$_2$ is lightly $n$-type. The diffusion current for the forward bias and the noise level of the saturation current for the reverse bias are observed, which clearly shows the rectified behavior. Hence, the band alignment is the staggered type II, as shown in **Figure 3**b. For the range of (ii) -8 V < $V_{TG}$ < 15 V (representative data; $V_{TG}$ = -7, 0 and 15 V), MoS$_2$ is highly *n*-type, and the band alignment become the staggered type II with larger band discontinuity. The reverse current is observed at $V_{TG}$ = -7 V. Since the thermally excited carrier through the $p^+$-*n* junction has already been suppressed at 20 K, this is the BTBT current from the valence band of $p^+$-WSe$_2$ to the conduction band of *n*-MoS$_2$. The further increase in $V_{TG}$ from -7 V to 15 V enhances the BTBT current. For the forward bias, the diffusion current for the $V_{TG}$ range (ii) is clearly observed, which is larger than that for the $V_{TG}$ range (i) because both electrons and holes flow through the $p^+/n^+$ junction. However, the NDR trend is not evident in this diode. **Figure 3**c shows the on-set voltage for BTBT as a function of $V_{TG}$, as shown by an arrow in **Figure 3**a, which indicates the valence band edge position of MoS$_2$ relative to the conduction band edge position of WSe$_2$. The valence band edge is effectively moved upward for $V_{TG}$ = -8 V to -6 V, whereas it is almost no change for $V_{TG}$ > -6 V. The reason is that $E_F$ in MoS$_2$ almost enters the conduction band edge, and further modulation of $E_F$ is limited because of the practically present interface states near the conduction band edge and large *DOS* in the conduction band. Therefore, even under the maximum $V_{TG}$ of 15 V, the band alignment for $p^+$-WSe$_2$/*n*-MoS$_2$ remains staggered type II. If the band alignment is broken type III, the diode properties with the NDR trend, which are schematically shown by the dark dotted line, should appear. For the present $p^+$-WSe$_2$/*n*-MoS$_2$ diode, the band alignment becomes broken type III only with the additional drain bias, which results in the observation of the BTBT current.

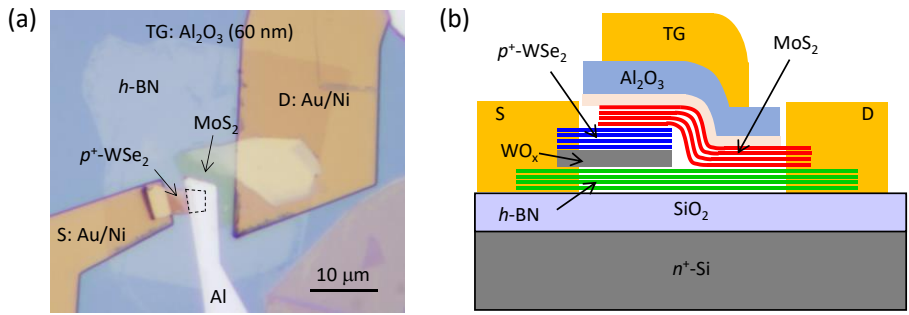

**Figure 2**. a) Optical micrograph and b) Schematic drawing of $p^+$-WSe$_2$/$n$-MoS$_2$ heterostructure TFET.

**Figure 3**d shows the temperature dependence of the diode properties at the maximum gate bias of $V_{TG}$ = 15 V, $V_{BG}$ = 0 V. For the reverse bias, an Arrhenius plot of the drain current at the reverse bias of -2 V is

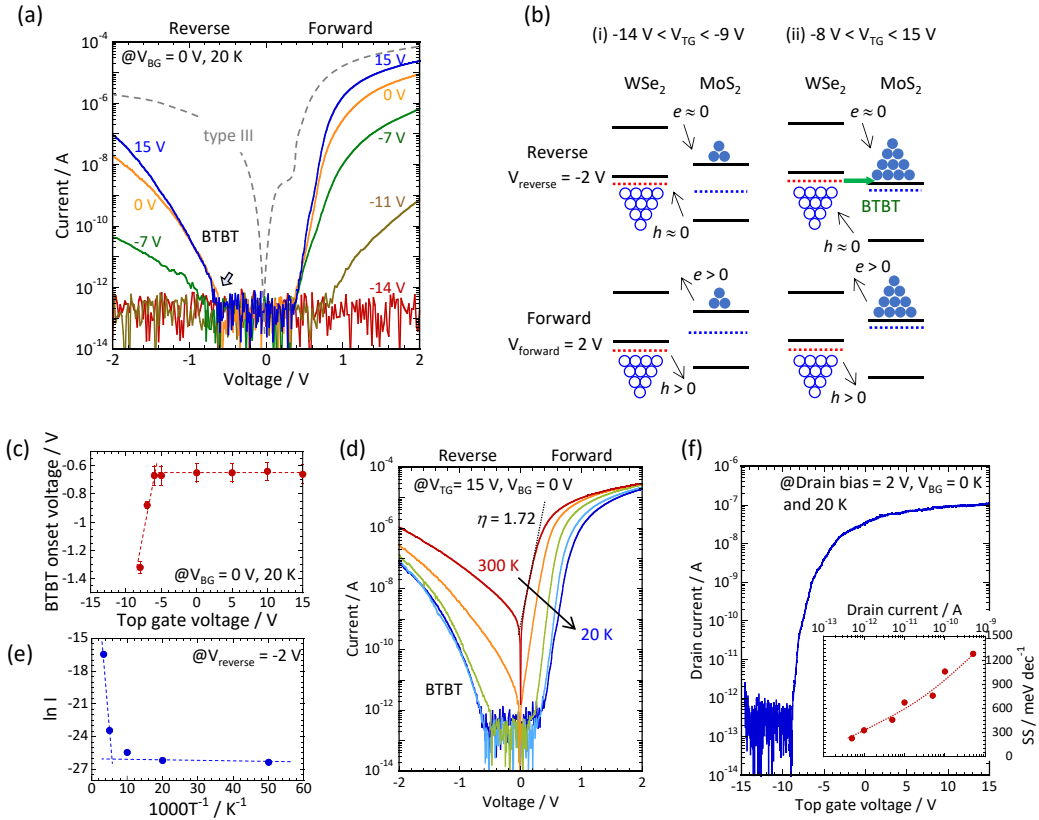

**Figure 3**. a) Diode properties in the $p^+$-WSe$_2$/*n*-MoS$_2$ heterostructure for different $V_{TG}$ at $V_{BG}$ = 0 V and 20 K. b) Schematic illustration of the band diagrams at forward and reverse biases for different $V_{TG}$ range at 20 K. $V_{reverse}$ and $V_{forward}$ indicate the reverse and forward voltages in a, respectively. c) BTBT onset voltage as a function of $V_{TG}$. d) Temperature dependence of the diode properties at $V_{TG}$ = 15 V and $V_{BG}$ = 0 V. The temperatures are 300 K, 200 K, 100 K and 50 K. The dotted line indicates the ideality factor of 1.72 at 300 K. e) Arrhenius plot of the drain current at the reverse bias of -2 V. f) Transfer characteristic of the device at the drain bias of 2 V and $V_{BG}$ = 0 V and at 20 K. The inset shows SS as a function of drain current.

conducted, as shown in **Figure 3**e. In the high-temperature region, the thermally activated behavior is observed, whereas the temperature-independent behavior is evident in the low-temperature region, which also supports the presence of the BTBT current. For the forward bias, the ideality factor $\eta$ for the diode is 1.72 at RT, which is superior to 3.46 for the WSe$_2$/SnSe$_2$ heterostructure diode.[16] This result suggests that the present $p^+$-WSe$_2$/$n$-MoS$_2$ interface is reasonably good because of the air stability of both 2D crystals.

**Figure 3**f shows the transfer characteristic of the device at the drain bias of 2 V, $V_{BG}$ = 0 V and 20 K. When the drain bias of 2 V is applied at $V_{TG}$ = 0 V and $V_{BG}$ = 0 V, the band alignment is broken type III. Therefore, this TFET exhibits the off state in the negative-$V_{TG}$ region, which can be modified by selecting the top-gate metal with the proper work function. The inset shows SS as a function of the drain current. The SS increases with increasing the drain current ($I_D$), i.e., the conduction band in the source overlaps with the valence band in the channel, which is consistent with the SS behavior in the TFET. The lowest SS value is 210 mVdec$^{-1}$, which is relatively higher than the target value of 60 mVdec$^{-1}$. Since the $p^+$-WSe$_2$/$n$-MoS$_2$ interface is reasonably good based on the ideality factor, the SS can be improved by improving the high-$k$/MoS$_2$ interface quality.[31] The enhancement of the top-gate capacitance relative to the interface trap capacitance by reducing the ALD Al$_2$O$_3$ thickness from 60 nm to ~10 nm is effective.

**2.3 $p^+$-WSe$_2$/$n$-WSe$_2$ type-III heterostructure**

In the $p^+$-WSe$_2$/$n$-MoS$_2$ diode, the band alignment under the maximum gate bias without the drain bias does not change from the staggered type II to the broken type III. Here, WSe$_2$ is used as a channel as well as $p^+$-source. The advantage of 2D crystals is the band gap tuning by selecting the layer number.[26] In the $p^+$-WSe$_2$/$n$-WSe$_2$ heterostructure TFET, the trilayer WSe$_2$ is selected as the channel because $E_G$ for the channel material (~1.3 eV for trilayer) should be larger than that for the source material (~1 eV for multilayer) to reduce the width of the energy barrier at the source junction.[1] The main difference from the device structure in **Figure 2**b is the thickness of the ALD-Al$_2$O$_3$ insulator (~30 nm), except the channel material. $p^+$-WSe$_2$ with the WO$_x$ layer and $h$-BN have almost identical thickness. The heterointerface area is ~33 $\mu$m$^2$. For the contact metal, Ni is used for both $p^+$-WSe$_2$ and $n$-WSe$_2$.

First, the transfer characteristics of the multilayer WSe$_2$ FET with the ~30-mm-thick ALD-Al$_2$O$_3$ top-gate insulator are investigated, as shown in **Figure S4** (Supporting Information). The $n$-channel behavior is clearly observed. By reducing the temperature, the drain current is drastically reduced, which is different from the case of the MoS$_2$ FET because the thermionic current through the Schottky barrier at the Ni/WSe$_2$ contact is reduced. Therefore, for the low-temperature measurement, $V_{BG}$ as well as $V_{TG}$ should be applied to recover the drain current.

Now, let us examine the diode properties of the $p^+$-WSe$_2$/$n$-WSe$_2$ heterostructure, as shown in **Figure 4**a, where the moderately low temperature of 150 K and $V_{BG}$ = 30 V are selected to minimize the effect of the $n$-WSe$_2$/Ni contact. For $V_{TG}$ = -8 and -6 V, the current is the noise floor for the entire voltage range because $n$-WSe$_2$ in the heterostructure region is in the off state. For $V_{TG}$ = -4 and -2 V, only the diffusion current is observed for the forward bias, whereas no reverse current is detected. When $V_{TG}$ increases from 0 to 8 V, the NDR trend is evident for the forward bias, whereas the BTBT current is clearly observed for the reverse bias. In addition to the NDR trend, the BTBT onset voltage is always zero, which indicates that the band alignment for the present $p^+$-WSe$_2$/$n$-WSe$_2$ heterostructure is the broken type III. The $V_{TG}$ dependence of the peak position for the NDR trend is not clear possibly because the valence band for the source is notably near the conduction band for the channel, and $E_F$ in the $n$-WSe$_2$ channel is already near the conduction band edge, although the band alignment is the broken type III, as shown in **Figure 4**c. To further support the presence of the NDR trend, the temperature dependence of the diode properties is shown at $V_{TG}$ = 2 V and $V_{BG}$ = 15 V in **Figure 4**b. The NDR trend diminishes after 260 K because of the thermionic current for the forward bias, which has been reported.[11] In general, the recombination current is apparent at the high temperature, which suggests that the observed depression in the forward bias is indeed the NDR trend. The temperature dependence of the BTBT current for the reverse bias results from the temperature dependence of the thermionic emission through the Ni/$n$-WSe$_2$ Schottky barrier.

**Figure 4**d shows the transfer characteristic of the device at different temperatures at the drain bias of 1 V and $V_{BG}$ = 20 V. **Figure 4**e shows the SS as a function of the drain current and temperature. The SS extracted at 50 K increases with increasing $I_D$, whereas the SS extracted at $I_D$ = 1×10$^{-11}$ A exhibits no temperature dependence. These results also prove the TFET operation. Although the lowest SS value is 210 mVdec$^{-1}$, the average SS for two decades from 10$^{-12}$ A to $10^{-10}$ A is ~400 mVdec$^{-1}$. The SS can be further improved by addressing the high-$k$/MoS$_2$ interface quality.

Finally, let us discuss the advantage of the present charge-transfer-type $p^+$-WSe$_2$ source. Some 2D crystals such as SnSe$_2$, BP and SnS naturally exhibit the defect-originated $p^+$ behavior. Although we have fabricated a $p^+$-SnS/$n$-SnS$_2$ heterostructure, the top-gate fabrication process degrades the diode properties. The prominent NDR has been reported in the BP/SnSe$_2$ heterostructure. However, an amorphous layer was observed at the heterointerface.[32] Therefore, the use of typical stable transition metal dichalcogenides as the source is an important step to realize 2D-2D TFETs.

### 3. Conclusion

The key finding in this study is the formation of the charge-transfer-type $p^+$-WSe$_2$ on $h$-BN substrates, which shows excellent process stability and negligible temperature dependence on the transfer characteristics. Using this $p^+$-WSe$_2$ as the source, two types of all solid-state 2D-2D heterostructure TFETs were fabricated: $p^+$-WSe$_2$/$n$-MoS$_2$ and $p^+$-WSe$_2$/$n$-WSe$_2$. The temperature dependence of the diode properties reveals that the band alignments for the former and latter are the staggered type II and broken type III under the gate bias condition, respectively. Therefore, the BTBT is demonstrated for both heterostructures, whereas the NDR trend is clearly observed for the $p^+$-WSe$_2$/$n$-WSe$_2$ heterostructure. This work suggests that high doped 2D crystal of the charge transfer type is an excellent choice as sources for TFETs.

### 4. Experimental section

MoS$_2$ natural bulk crystal is purchased from SPI Supplies, whereas WSe$_2$ bulk crystal is grown using a physical vapor transport technique without the I$_2$ transport agent.[33] All 2D thin layers were prepared from the mechanical exfoliation of these bulk crystals. The $p^+$-WSe$_2$/$n$-MoS$_2$ heterostructure on the $h$-BN substrate was fabricated using a dry transfer method with PDMS in the alignment system.[27] No additional annealing was performed after the heterostructure fabrication. Ni/Au was deposited as the source/drain electrodes after the electrode pattern formation using electron beam lithography. Then, Y metal with a thickness of 1 nm was deposited via thermal evaporation of the Y metal from a PBN crucible in an Ar atmosphere with a partial pressure of $10^{-1}$ Pa; afterwards, oxidization in the laboratory atmosphere was performed to form the buffer layer.[28,29] The $Al_2O_3$ oxide layer with a thickness of ~60 nm was deposited via ALD at 200°C, followed by the Al top-gate electrode formation. Alternatively, the fabrication procedure of the $p^+$-WSe$_2$/$n$-WSe$_2$ heterostructure on the *h*-BN substrate is identical, except the ALD-$Al_2O_3$ layer was ~30 nm thick.

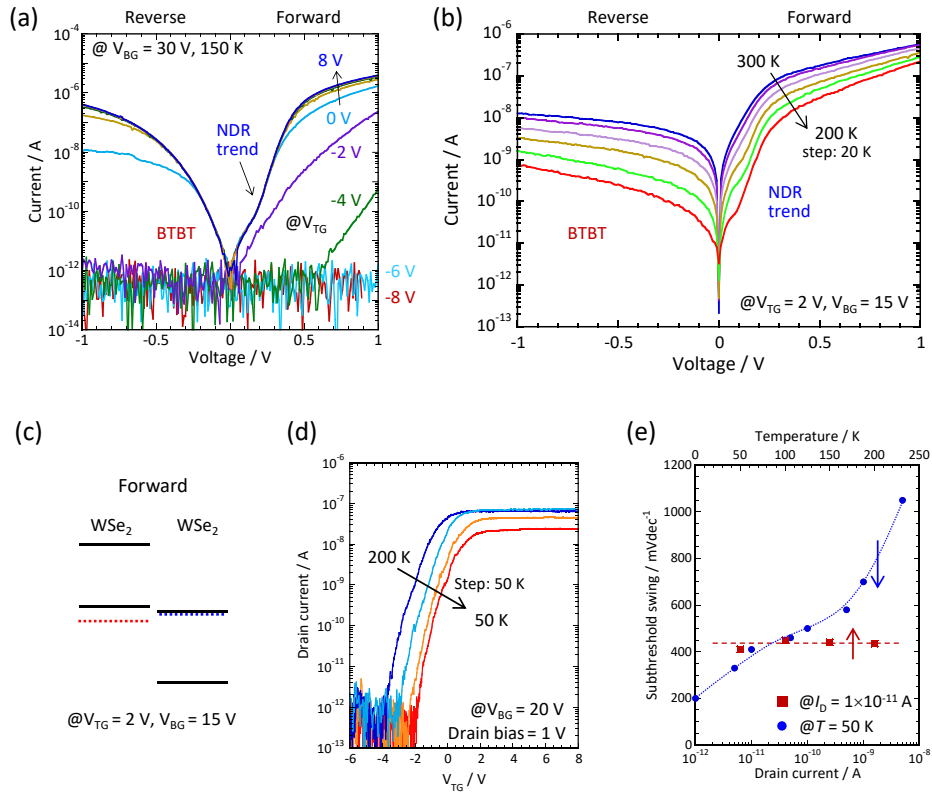

**Figure 4.** a) Diode properties in the $p^+$-WSe$_2$/$n$-WSe$_2$ heterostructure at different $V_{TG}$ at $V_{BG}$ = 30 V and 150 K. b) Diode properties in the $p^+$-WSe$_2$/$n$-WSe$_2$ heterostructure at different temperatures at $V_{TG}$ = 2 V and $V_{BG}$ = 0 V. c) Schematic illustration of the band diagrams at forward and bias for the $V_{TG}$ range from 0 V to 8 V at 150 K. d) Transfer characteristics of the device at different temperatures at the drain bias of 1 V and $V_{BG}$ = 20 V. e) SS as a function of the drain current and the temperature.

The layer number of 2D flakes was determined using Raman microscopy with a 488 nm Ar laser and photoluminescence spectroscopy. The thickness of the 2D flakes was determined by correlating the AFM thickness and the optical contrast on the substrate. All electrical measurements were performed in a vacuum prober with a cryogenic system.

**Supporting Information**
Supporting Information is available from the Wiley Online Library or from the authors.

**Acknowledgements**
This research was supported by the JSPS Core-to-Core Program, A. Advanced Research Networks, JSPS KAKENHI Grant Numbers JP25107004, JP16H04343, JP16K14446, and JP26886003, and JST PRESTO Grant Number JPMJPR1425, Japan.

**Conflict of Interest**
The authors declare no conflict of interest.

**Keywords:**
2 dimensional heterostructure, Band to band tunneling, Negative differential resistance, Subthreshold swing

**Supporting Information**

**2-Dimensional Tunnel FETs with a Stable Charge-Transfer-Type p$^+$-WSe$_2$ Source**

*Junyang He, Nan Fang, Keigo Nakamura, Keiji Ueno, Takashi Taniguchi, Kenji Watanabe, and Kosuke Nagashio*$^*$

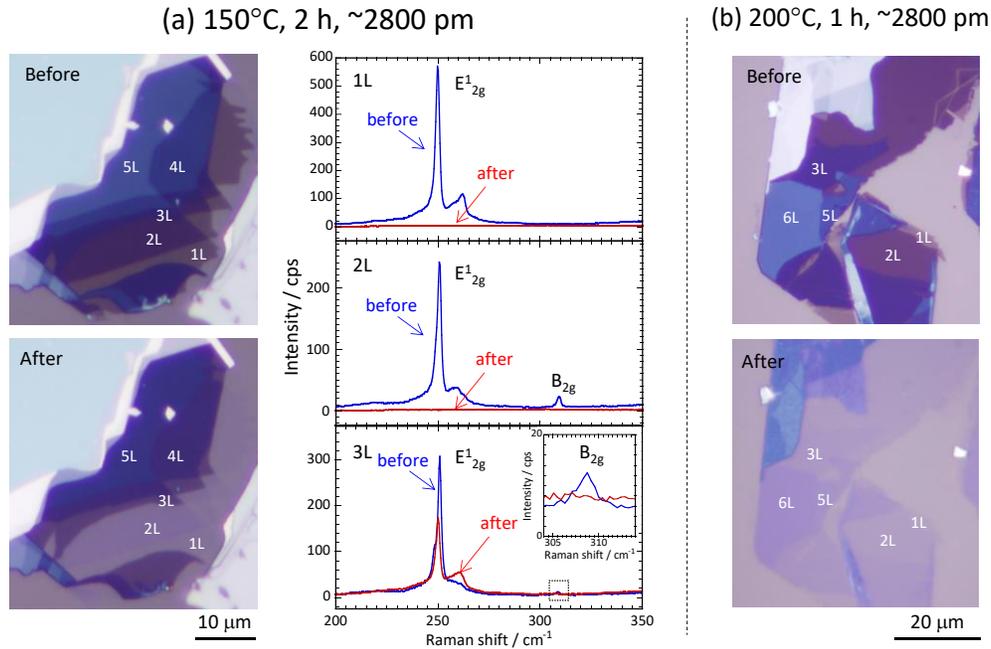

**Figure S1.** In this study, O$_3$ is produced by a commercial dielectric barrier discharge ozone generator. The concentration of O$_3$ just at the outlet is ~2800 ppm, which is the minimum concentration in this system. O$_3$ generally becomes inactive rapidly. Therefore, the generated O$_3$ concentration is considerably reduced through the homemade gas line from ozone generator to the rapid thermal annealing system (RTA). Therefore, the O$_3$ annealing conditions for WO$_x$ formation on WSe$_2$ were investigated. (a) At the conditions of 150°C, 2 hours, and ~2800 ppm O$_3$, 2 layers are oxidized, which is supported from the following results. After the annealing, the Raman signals indicating 2 L disappeared, and Raman spectra for 3 L became similar to those of 1 L without B$_{2g}$ [a]. Moreover, the 2 L became transparent because of WOx formation, and the optical contrast of the 3 L became similar to that for 1 L. These results are consistent with the previous report [b]. (b) At the conditions of 200°C, 1 hour, and ~2800 ppm O$_3$, 6 layers are oxidized. ~6 L became transparent. In terms of the source in TFET, the bulk WSe$_2$ with a large density of states and constant band gap is proper. Therefore, in this study, the annealing conditions of 200°C, 1 hour, and ~2800 ppm O$_3$ were applied to the bulk WSe$_2$ with initial thickness of ~15 nm.

[a] X. Luo, Y. Zhao, J. Zhang, M. Toh, C. Kloc, Q. Ziong, S. Y. Quek, Effects of lower symmetry and dimensionality on Raman Spectra in two-dimensional WSe$_2$, *Phys. Rev. B*, **2013**, *88*, 195313.
[b] M. Yamamoto, S. Dutta, S. Aikawa, S. Nakaharai, K. Wakabayashi, M. S. Fuhrer, K. Ueno, K. Tsukagoshi, Self-limiting layer-by-layer oxidation of atomically thin WSe$_2$, *Nano lett.*, **2015**, *15*, 2067.

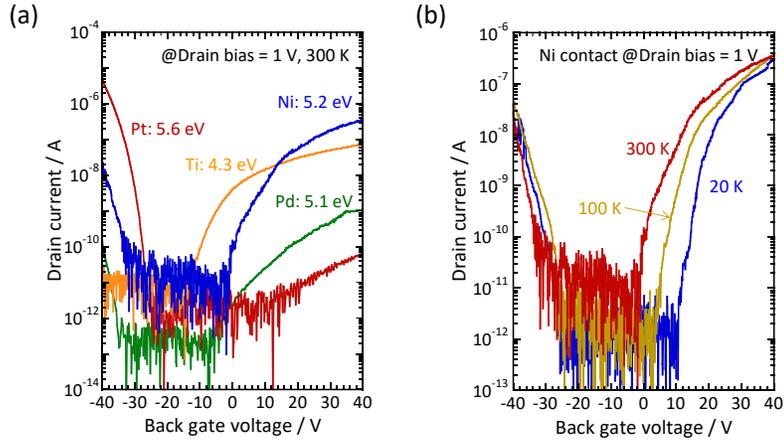

**Figure S2.** a) Transfer characteristics for back gate monolayer WSe$_2$ FET on SiO$_2$(90 nm)/$n^+$-Si with different contact metals, Ti, Pd, Ni, and Pt. Ni shows both *n* and *p* type behavior. When the thickness of WSe$_2$ is increased from monolayer to bulk, the band gap is reduced from ~1.5 eV to 1 eV. For bulk WSe$_2$, Ni works as better bipolar electrode. b) Temperature dependence of transfer characteristics for back gate monolayer WSe$_2$ FET. Because of the carrier modulation in WSe$_2$ just below the contact metal, the drain current does not reduce at high gate bias region, although the threshold voltage shift is clearly observed.

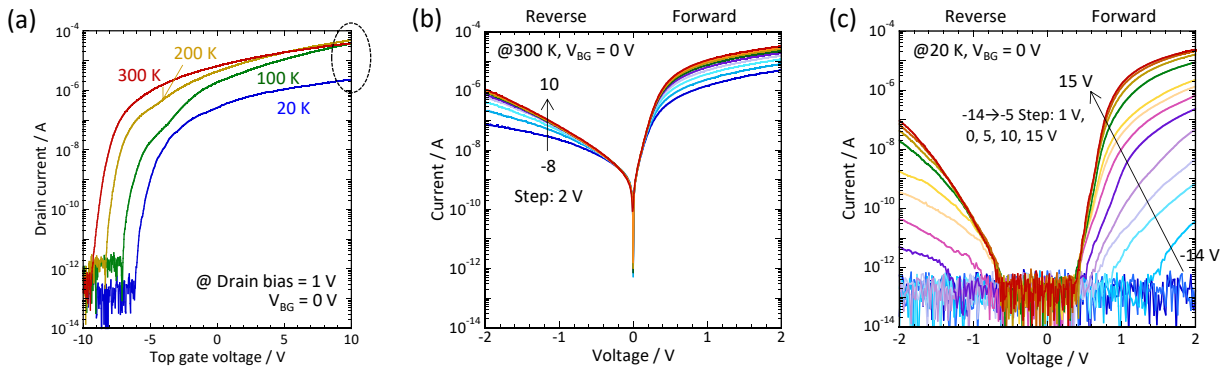

**Figure S3.** a) Transfer characteristics of multilayer MoS$_2$ FET with the Al$_2$O$_3$ top gate (~60 nm), which is fabricated at the different position on the same SiO$_2$/Si wafer. Diode properties for $p^+$-WSe$_2$/$n$-MoS$_2$ heterostructure at different $V_{TG}$ at the temperatures of b) 300 K and c) 20 K.

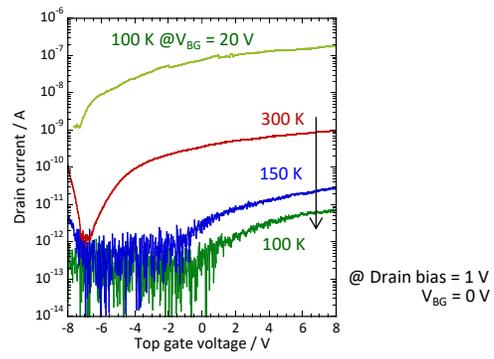

**Figure S4.** a) Transfer characteristics of multilayer $WSe_2$ FET with the $Al_2O_3$ top gate (~30 nm), which is fabricated at the different position on the same $SiO_2$/Si wafer.